\documentclass[10pt,letterpaper,twocolumn]{article} 

\usepackage{ol2}
\usepackage[draft,implicit=false]{hyperref}
\usepackage{amsmath}
\usepackage{balance}

\begin{document}

\twocolumn[ 

\title{Carrier envelope phase dynamics of cavity solitons: soliton stability and scaling law}

\author{Chengying Bao, Changxi Yang$^{*}$}

\address{State Key Laboratory of Precision Measurement Technology and Instruments, Department of Precision Instruments, Tsinghua University, Beijing 100084, China\\
$^*$Corresponding author: cxyang@tsinghua.edu.cn
}

\begin{abstract}
The relationship between carrier envelope phase (CEP) slip of cavity soliton (CS) and pump phase detuning is derived analytically and numerically. To preserve the stability of CS, CEP slip always equals to the pump phase detuning. When CEP slip fails to follow the pump phase detuning, CS becomes unstable. The locking between CEP slip and pump phase detuning results in a scaling law for CS.
\end{abstract}


 ] 

Kerr frequency combs (KFCs) generated from the parametric process in continuous wave (CW) laser pumped microresonators have been heavily investigated experimentally and theoretically in recent years \cite{Kippenberg_Science2011Rev,Kippenberg_Nat2007comb,Maleki_PRL2008tunable,Gaeta_NP2010cmos,Morandotti_NP2010cmos,
Weiner_Optica2014overmode,Diddams_NC2015phase,Wong_PRL2015normal,Tang_Optica2014Green,Chembo_PRA2013WGM,Coen_OL2013modeling}. It shows great potential in advancing applications such as arbitrary waveform generation \cite{Weiner_NP2011shaping}, compact optical atomic clock \cite{Diddams_Optica2014clock}, coherent data transmission \cite{Kippenebrg_NP2014coherent}, mid-infrared frequency comb generation \cite{Kippenberg_NC2013mid,Gaeta_NC2015mid,Wabnitz_OL2014Mid}. For numerous applications including optical atomic clock, high accuracy gas spectroscopy in the mid-infrared, the full stabilization of KFCs is highly desired. Del'Haye et. al. demonstrated that KFCs can be locked to a femtosecond frequency comb, by controlling the pump power and wavelength \cite{Kippenberg_PRL2008full}. However, self-referencing stabilization of KFCs is more favorable. The excessive noise in the previously reported octave-spanning comb hindered the self-referencing detection \cite{Kippenberg_PRL2011Octave,Gaeta_OL2011octave}. Recently, it has been found that a KFC undergoes transition from the chaotic regime into the highly coherent phase-locking regime, by tuning the pump phase detuning at certain speed \cite{Gaeta_OL2013route,Kippenberg_NP2014Soliton}. The observation of cavity soliton (CS) in microresonators \cite{Gaeta_OE2013modelocking,Kippenberg_NP2014Soliton} sheds light on the self-referencing stabilization of KFCs. Exploiting external spectrum broadening of CS in a highly nonlinear fiber, Jost et. al. managed to measure the offset frequency signal ($f_{0}$) of a KFC by 2f-3f detection, with assistance from two auxiliary lasers locked to the KFC  \cite{Kippenberg_arXiv2014microwave}. To fulfill CS' prospects in ultrafast science, a lot of theoretical and experimental work have been done to unveil the properties of CS, ranging from pulse stability and pattern \cite{Chembo_PRA2014stablity,Coen_PRA2014dynamics,Maleki_OL2012breather}, scaling law \cite{Coen_OL2013universal}, soliton formation and mode-locking onset \cite{Kippenberg_PRL2014mode,Gaeta_arXiv2014self}, to phase-matching \cite{Bao_JOSAB2014mode} etc. However, most of the work focuses on the envelope of CS and little attention has been paid to the carrier envelope phase (CEP) dynamics of CS. Gaining more comprehension about the CEP dynamics for CS is needed for the self-referencing stabilization of KFCs. CS is strikingly different from the widely studied mode-locked lasers \cite{Cundiff_RMP2003colloquium}. For example, the CW pump itself is a line of the generated KFC and interferes with the CS coherently every roundtrip. Therefore, an investigation of CS' CEP dynamics will also improve our understanding of CS. In this Letter, we show that this coherent interference determines the CEP slip of CS, leading to a locking between it and the pump phase detuning. Our analyses of the CEP dynamics also give an intuitive explanation to the excitation of breather soliton (BS) \cite{Maleki_OL2012breather,Bao_JOSAB2014mode} and the scaling law on the peak power of CS.

In KFC generation, a CW pump is coupled into a high-Q microresonator and experiences strong enhancement in the high-Q resonator. The high optical intensity stored in the microresonator will generate numbers of evenly spaced comb lines via cascaded four-wave-mixing (CFWM) (Fig 1(a)) \cite{Kippenberg_Science2011Rev}. Under an appropriate pump condition, the comb lines can be phase locked and manifest as ultrashort CS in the time domain. As a feature of CS, it will interfere with the CW pump every roundtrip, when it passes through the coupling region between the microresonator and the waveguide (or tapered fiber). Then, phase delay between the CS and the CW pump determines output of the interference. To stay stable, the phase delay between them should be kept the same for different roundtrips. For the CW pump, the phase delay between two successive interference ($\Delta\phi_{cw}$) is determined by the pump frequency and roundtrip time, i.e., $\Delta\phi_{cw}=2N\pi-\omega_{cw}\text{T}_{r}$, where $\omega_{cw}$ is the angular frequency of the CW pump, $N$ is a known integer and $\text{T}_{r}$ is the roundtrip time of the cavity, which can be measured directly, when the repetition rate $f_{r}$ is smaller than 100 GHz. For the CS, the phase delay for two adjacent CSs is the CEP slip ($\Delta\phi_{CEP}$) accumulated during propagation in the microresonator, which has contribution from both linear and nonlinear effects \cite{Haus_OL2001group,Cundiff_OL2004CEP}. An illustration of such a process can be found in Fig. 1 (b). Therefore, for the requirement of the stability of CS, the following relationship should be satisfied,
\begin{equation}
 \Delta\phi_{CEP}=\Delta\phi_{cw},
\label{eq:CEPCW}
\end{equation}

\begin{figure}[t]
\includegraphics[width=\columnwidth]{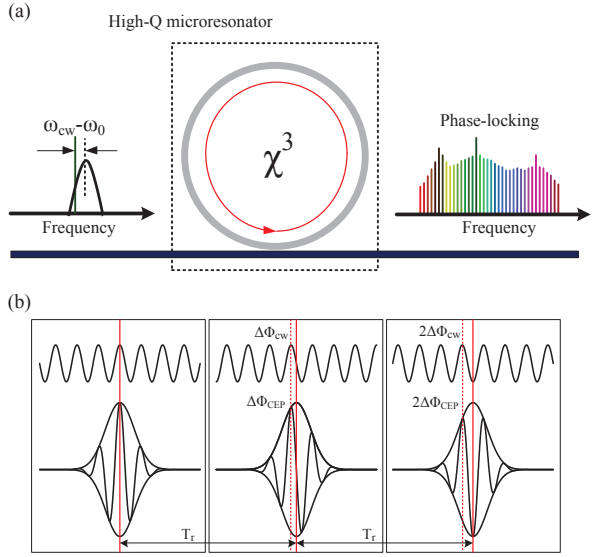}
\caption{(Color online) (a) Generation of KFC in a high-Q microresonator, phase-locking of the comb lines can be achieved by choosing appropriate pump phase detuning. (b) The illustration of the interference for three successive CSs with the CW pump. The red solid lines show the position of the pulse envelope peak, which determines the roundtrip time T$_{r}$, while the red dashed lines present the adjacent peak of the pulse-carrier next to the envelope peak and peak of the CW pump.}
\label{Fig:Illustration}
\end{figure}

The above relationship in Eq. \ref{eq:CEPCW} can also be derived analytically. Note that the pump is a line of the generated KFC. Thus, the offset frequency $f_{0}$ should be,

\begin{equation}
   f_{0}=f_{cw}-Nf_{r},
\label{eq:offsetfrequency}
\end{equation}
where $f_{cw}$=$\omega_{cw}/2\pi$. Recalling the relationship between the CEP slip and offset frequency $f_{0}$ \cite{Cundiff_RMP2003colloquium},

\begin{equation}
   f_{0}=-\frac{\Delta\phi_{CEP}}{2\pi}f_{r},
\label{eq:CEPoffset}
\end{equation}

Combining Eq. \ref{eq:offsetfrequency} and Eq. \ref{eq:CEPoffset}, we can get,

\begin{equation}
   \Delta\phi_{CEP}=-2\pi(f_{cw}-Nf_{r})/f_{r}=2N\pi-\omega_{cw}\text{T}_{r},
\label{eq:phase2}
\end{equation}
Thus, the relationship Eq. \ref{eq:CEPCW} can be derived, using the definition of $\Delta\phi_{cw}$.

To verify the lock between CEP slip and pump phase detuning, we carried out numerical simulation to examine the CEP slip. Here, we divide the evolution of the CS into two parts, i.e., propagation in the microresonator and interference with the CW pump \cite{Coen_OL2013modeling,Wabinitz_OC1992dissipative} to explicitly show the pump phase detuning and CEP slip in every roundtrip. Propagation in the resonator is governed by the generalized nonlinear schr\"{o}dinger equation (NLSE) as,

\begin{equation}
\begin{aligned}
\frac{\partial E_{n}(\tau,z)}{\partial z}&=-\frac{\alpha_{0}}{2} E_{n}(\tau,z)-i\frac{\beta_{2}}{2}\frac{\partial^{2}E_{n}}{\partial \tau^{2}}+i\gamma|E_{n}|^{2}E_{n}&\\&-\frac{\gamma}{\omega_{0}}\frac{\partial(|E_{n}|^{2}E_{n})}{\partial\tau},&
\end{aligned}
\label{eq:NLSE}
\end{equation}
where E$_{n}$($\tau$,z) is the envelope of the CS for the $n$th roundtrip propagation, and $\alpha_{0}$ is the intrinsic loss of the waveguide, $\beta_{2}$ is the group velocity dispersion (GVD), $\gamma$ is the nonlinear coefficient of the cavity. The Kerr shock term is included in the equation, as it is important in determining the group velocity of solitons \cite{Haus_OL2001group}.
The interference of the CS with the CW pump is described as,

\begin{equation}
\begin{aligned}
E_{n+1}(\tau,z=0)&=\sqrt{1-T^2}E_{n}(\tau,z=L)&\\&+TE_{in}\text{exp}[i(n\Delta\phi_{cw,m}+\phi_{acc,m})],&
\end{aligned}
\label{eq:Coupling}
\end{equation}
where $E_{in}$ is the amplitude of the CW pump, T is the coupling coefficient and $\Delta\phi_{cw,m}$ is the pump phase detuning. As we tune the pump phase detuning to find the mode-locking state \cite{Gaeta_OL2013route}, the subscript $m$ is used to denote the $m$th chosen pump phase detuning. In addition, we use $\phi_{acc,m}=\sum_{p=1}^{m-1}n_{p}\Delta\phi_{cw,p}$, where $n_{p}$ is the number of roundtrips run for the $p$th pump phase detuning, to stand for the pump phase delay accumulated before tuning the pump phase detuning to $\Delta\phi_{cw,m}$. We choose T=0.1, $\beta_{2}$=-220 ps$^{2}$/km, $\alpha_{0}$=0.15 dB/cm, $\gamma$=0.8 W$^{-1}$m$^{-1}$, $\omega_{0}$=2$\pi\ast$193 THz and run the simulation with white noise as initial condition until convergence of $E_{n}(\tau,z)$.

\begin{figure}[t]
\includegraphics[width=0.95\columnwidth]{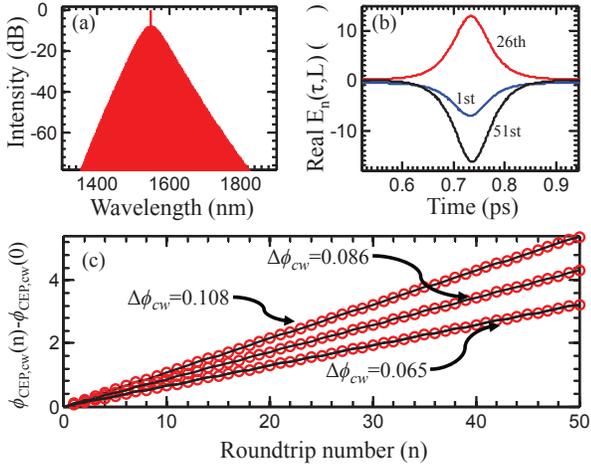}
\caption{(Color online) (a) Spectrum of the generated CS, (b) real part of the CS' electric field at different roundtrips, (c) CEP (red circles) and accumulated pump phase detuning (black lines) for different roundtrips under different $\Delta\phi_{cw}$. CEP always follows the change of pump phase detuning under different $\Delta\phi_{cw}$.}
\label{Fig:relation}
\end{figure}

At the pump power of 0.59 W, CS can be excited when increasing $\Delta\phi_{cw,m}$ slowly from 0.016 to 0.065. CS remain stable when further increasing $\Delta\phi_{cw}$ to 0.108 and the corresponding spectrum of the intracavity CS is shown in Fig. 2(a). The generated comb has a typical sech$^{2}$-shape, i.e., triangle-shape in logarithmic scale \cite{Kippenberg_NP2014Soliton,Coen_OL2013universal}. Although the carrier frequency is assumed to be zero in NLSE, the real part of the simulated electric field changes with roundtrip owing to the CEP slip (Fig. \ref{Fig:relation}(b)). The carrier of the CS can be reconstructed by multiplying $exp(-i\omega_{0}t)$ to $E_{n}$, where $t$ is the retarded time frame, with $t=0$ being the peak of CS' envelope \cite{Cundiff_RMP2003colloquium}. We extract the CEP of the CS by getting the phase of the optical field at the peak of the CS' envelope, where the $exp(-i\omega_{0}t)$ term equals to 1. The extracted CEP of the pulse, shown in Fig. 2(c), increases linearly with the roundtrip number. Furthermore, the linear change always follows the accumulation of the pump phase detuning, for different $\Delta\phi_{cw}$ as long as the CS remains stable.

\begin{figure}[t]
\includegraphics[width=\columnwidth]{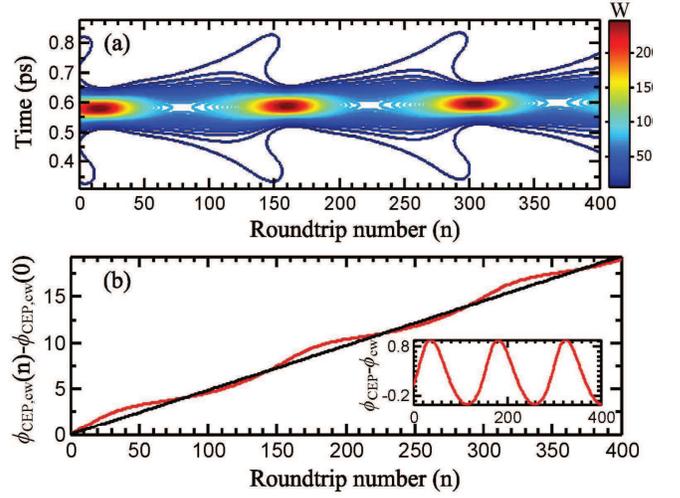}
\caption{(Color online) (a) Temporal evolution of breather soliton excited at the pump power of 0.59 W, and $\Delta\phi_{cw}$=0.049. (b) The CEP of the breather soliton (black line) and the accumulated pump phase detuning (red line) at different roundtrips. The inset shows the phase differential between them.}
\label{Fig:breather}
\end{figure}

CS is only stable in a certain pump phase detuning range and becomes unstable (also referred as BS), when the pump phase detuning decreases to a certain value \cite{Coen_OL2013universal}. Here, we find that BS will be excited when decreasing the pump phase detuning to 0.049. The BS stretches and compresses periodically in $\sim$140 roundtrips (Fig. \ref{Fig:breather}(a)). For the BS, the CEP no longer changes linearly and deviates from the pump phase detuning accumulation. The difference between the CEP and the accumulated $\Delta\phi_{cw}(n)$, which is depicted in the inset of Fig. \ref{Fig:breather}(b), also changes periodically in $\sim$140 roundtrips. Consequently, the soliton changes its shape and peak power slightly every time it passes through the coupling region. The slight change is amplified when it propagates in the microresonator due to the unbalance of dispersion and nonlinearity. As a result of the periodically changing phase delay between them, the soliton exhibits the periodic variation in the time domain.

In addition to affecting the stability of CS, the relationship in Eq. \ref{eq:CEPCW} gives a restriction on the peak power of CS. Recalling the group velocity of soliton, the CEP slip of a optical soliton can be calculated as \cite{Haus_OL2001group},

\begin{equation}
   \Delta\phi_{CEP}=\frac{1}{2}\gamma|A_{0}|^{2}L-\beta_{0}^{\omega_{c}}L+\omega_{c}\beta_{1}^{\omega_{c}}L,
\label{eq:hausslip}
\end{equation}
where, A$_{0}$ is the amplitude of the soliton, L is the length of the cavity, $\beta_{0}^{\omega_{c}}$, $\beta_{1}^{\omega_{c}}$ are the propagation constant and the first-order dispersion at the carrier frequency $\omega_{c}$ respectively. Thus, $\omega_{c}$$\beta_{1}^{\omega_{c}}$L-$\beta_{0}^{\omega_{c}}$L is the linear CEP slip.

From Eq. \ref{eq:hausslip} and Eq. \ref{eq:CEPCW}, we can get a scaling law for the peak power of CS,

\begin{equation}
   |A_{0}|^{2}=2\frac{\Delta\phi_{cw}+\beta_{0}^{\omega_{c}}L-\omega_{c}\beta_{1}^{\omega_{c}}L}{\gamma L},
\label{eq:peak}
\end{equation}
so the peak power of the soliton is independent from the pump power and increases linearly with the pump phase detuning. It is worth mentioning the pump phase detuning ($\Delta\phi_{cw}$) defined here is different from the pump phase detuning in Lugiato-Lefever Equation (LLE), where the pump phase detuning is defined as $\Delta\phi_{cw}^{lle}$=-($\omega_{cw}-\omega_{r}$)T$_{r}$ \cite{Chembo_PRA2013WGM,Coen_OL2013universal}, with $\omega_{r}$ being the resonance of the pumped cavity mode. However, mod$_{2\pi}$($\omega_{r}$T$_{r}$) is not zero due to the difference in linear phase velocity (determines $\omega_{r}$) and group velocity (determines T$_{r}$). Hence, there is a difference between $\Delta\phi_{cw}$ and $\Delta\phi_{cw}^{lle}$,

\begin{equation}
   \Delta\phi_{cw}^{lle}-\Delta\phi_{cw}=\omega_{r}\text{T}_{r}-2N\pi=\omega_{r}\beta_{1}^{\omega_{c}}L-\beta_{0}^{\omega_{r}}L,
\label{eq:diff}
\end{equation}
where the last step is obtained by using the facts that $\beta_{0}^{\omega_{r}}L$=2N$\pi$ for the cavity resonance and T$_{r}$=$\beta_{1}^{\omega_{c}}$L.

By substituting $\Delta\phi_{cw}$ with $\Delta\phi_{cw}^{lle}$ in Eq. \ref{eq:peak}, we can get,

\begin{equation}
\begin{aligned}
   &|A_{0}|^{2}=2\frac{\Delta\phi_{cw}^{lle}-\omega_{r}\beta_{1}^{\omega_{c}}L+\beta_{0}^{\omega_{r}}L+\beta_{0}^{\omega_{c}}L-\omega_{c}\beta_{1}^{\omega_{c}}L}{\gamma L}&\\&\simeq2\frac{\Delta\phi_{cw}^{lle}-\beta_{1}^{\omega_{c}}(\omega_{c}+\omega_{r})L+2\beta_{0}^{\omega_{c}}L+
   \beta_{1}^{\omega_{c}}(\omega_{r}-\omega_{c})L}{\gamma L}&\\
   &\simeq2\frac{\Delta\phi_{cw}^{lle}+2(\beta_{0}^{\omega_{c}}L-\omega_{c}\beta_{1}^{\omega_{c}}L)}{\gamma L}.&
\label{eq:llepeak}
\end{aligned}
\end{equation}
This scaling law is consistent but different from the scaling law for CS' peak power derived via LLE model in \cite{Coen_OL2013universal}, where the normalized peak power of the CS is found to be twice of the normalized pump phase detuning. The scaling law in \cite{Coen_OL2013universal} can be ascribed to the analytical solution to LLE under pulsed pumping \cite{Smirnov_PRE1996existence}. The solution is used to describe to the envelope of CS without considering the CEP dynamics of CS. However, from Eq. \ref{eq:llepeak}, we can see the linear CEP slip plays an important role in determining the property of CS and pumping in the band where $\beta_{0}^{\omega_{c}}L-\omega_{c}\beta_{1}^{\omega_{c}}L$ is positive is beneficial for getting higher peak power and shorter CS.

In conclusion, the relationship between CEP slip and pump phase detuning is derived for CS. To prevent the coherent interference with the CW pump from changing CS' property, CEP slip of CS should be equal to the pump phase detuning. Otherwise, BS can be excited. Based on the CEP dynamics, a modified scaling law of CS is further obtained with the impact from the linear CEP slip included. These results give important insights on CS and could permit a novel method for the detection of $f_{0}$ signal by measuring the pump phase detuning without the need of octave-spanning spectrum.

~~~~

We acknowledge support from Natural Science Foundation of China (61177046, 61377039).
\balance

\end{document}